\documentclass[prd,aps,nofootinbib,preprintnumbers]{revtex4}
\usepackage{amssymb}
\usepackage{bm}
\usepackage{enumerate}
\usepackage{amssymb}
\usepackage{amsmath}
\usepackage{feynmf}
\usepackage{slashed}
\usepackage{graphicx}
\usepackage{filecontents}

\newcommand\be{\begin{equation}}
\newcommand\ee{\end{equation}}

\newcommand\bea{\begin{eqnarray}}
\newcommand\eea{\end{eqnarray}}

\begin{document}

\title
{\Large \bf Non-thermal Production of Dark Matter from Primordial Black Holes}

\author{Rouzbeh Allahverdi$^{1}$}
\author{James Dent$^{2}$}
\author{Jacek Osinski$^{1}$}

\affiliation{$^{1}$~Department of Physics and Astronomy, University of New Mexico, Albuquerque, NM 87131, USA}
\affiliation{$^{2}$~Department of Physics, Sam Houston State University, Huntsville, TX 77341, USA}

\begin{abstract}
We present a scenario for non-thermal production of dark matter from evaporation of primordial black holes. A period of very early matter domination leads to formation of black holes with a maximum mass of $\simeq 2 \times 10^8$ g, whose subsequent evaporation prior to big bang nucleosynthesis can produce all of the dark matter in the universe. We show that the correct relic abundance can be obtained in this way for thermally underproduced dark matter in the 100 GeV-10 TeV mass range. To achieve this, the scalar power spectrum at small scales relevant for black hole formation should be enhanced by a factor of ${\cal O}(10^5)$ relative to the scales accessible by the cosmic microwave background experiments. 
%This enhancement is compatible with Planck data and can in principle be achieved via ultra slow-roll motion toward the end of inflation.         
\end{abstract}
\maketitle

\section{Introduction}

Various lines of evidence support the existence of dark matter (DM) in the universe~\cite{BHS}. Weakly interacting massive particles (WIMPs) are promising candidates for DM in the Universe, and major direct, indirect, and collider searches are currently underway to discover WIMP-like DM and determine its properties. The relic abundance of WIMPs may be nicely explained via the well-explored ``WIMP miracle". This paradigm assumes the universe was in a radiation-dominated (RD) phase at temperatures about the DM mass $m_{{\rm{DM}}}$. The DM relic abundance in this picture is set when the annihilation rate of DM particles drops below the Hubble expansion rate, a process called ``thermal freeze-out", which typically happens at a temperature $T_f \sim m_{{\rm{DM}}}/20$. The observed DM abundance is obtained if the annihilation rate takes the value $\langle \sigma_{{\rm{ann}}} v \rangle_f = 2-3 \times 10^{-26}$ cm$^3$ s$^{-1}$, nearly independently of the dark matter mass above values of $m_{\chi} \approx 10$ GeV (and variation with mass roughly up to a factor of two for masses below 10 GeV)  \cite{Steigman:2012nb}. 

The WIMP miracle has been a staple of DM physics for decades, but is coming under increasing pressure in light of recent experimental data. Notably, Fermi-LAT's results from observations of dwarf spheroidal galaxies~\cite{fermi2015} and newly discovered Milky Way satellites~\cite{fermi2016} rule out the nominal value of the annihilation rate in various channels for DM masses up to $\sim 100$ GeV. Moreover, a RD phase at the time of freeze-out is an \emph{assumption}, as we currently have no direct observational probe of the early universe prior to the onset of big bang nulceosynthesis (BBN). A non-standard thermal history is a generic feature of various early universe models including some string theory constructions (for a review, see~\cite{KSW}). The late decay of modulus fields, which can dominate the universe's energy density leading to an epoch of early matter domination (EMD), reheats the universe to temperatures below $T_f$, thereby rendering the WIMP miracle irrelevant in this framework. 

%	author		=  {{Poulin}, Vivian and {Serpico}, Pasquale D. and {Calore}, Francesca, and {Clesse}, Sebastien and {Kohri}, Kazunori},

Exploring alternatives to the WIMP miracle is therefore well motivated by both theoretical and experimental considerations, and has been gaining increased attention. An attractive scenario is non-thermal production of DM from late decay of moduli that drive an EMD era. This scenario can accommodate both cases with large annihilation rate $\langle \sigma_{{\rm{ann}}} v \rangle_f > 3 \times 10^{-26}$ cm$^3$ s$^{-1}$ (leading to thermal underproduction)~\cite{MR,Scott} and small annihilation rate $\langle \sigma_{{\rm{ann}}} v \rangle_f < 3 \times 10^{-26}$ cm$^3$ s$^{-1}$ (resulting in thermal overproduction)~\cite{GG,branching} (for explicit examples in the context of type IIB string compactifications, see~\cite{ACDS}).

Also, as WIMP detection remains elusive, alternative models in which DM is not a WIMP-like particle have attracted significant attention in recent years (for example, see~\cite{Battaglieri:2017aum}). One alternative is that DM, instead of being an elementary particle, is (at least partially) composed of primordial black holes (PBHs)~\cite{Carr:1974nx,Carr:1975qj} formed in the early universe.\footnote{For other early works as well as some reviews, see~\cite{Khlopov:1985jw,Polnarev:1986bi,Carr:2005zd,Khlopov:2008qy,Carr:2009jm,Carr:2016drx,Garcia-Bellido:2017fdg}.} PBHs with a mass $\gtrsim 10^{15}$g would not have evaporated by the present time, and may in principle constitute a fraction or, perhaps, all of the DM in the universe. While this possibility has been a subject of study for a long time, the recent discovery of gravitational waves (GW) by the Advanced LIGO group has led to intensified efforts to constrain a possible PBH population.\footnote{This includes examination of the LIGO measurement and constraints on the PBH mass distribution (including the possibility of extended mass distributions) and merger rate \cite{Bird:2016dcv,Sasaki:2016jop,Clesse:2016vqa,Kovetz:2016kpi,Green:2016xgy,Kuhnel:2017pwq,Carr:2017jsz,Bellomo:2017zsr,Inomata:2017bwi}, general signatures of exotic compact objects \cite{Giudice:2016zpa}, possible PBH progenitors \cite{Raccanelli:2016cud,Postnov:2017itw}, future GW searches and expectations \cite{Kyutoku:2016ppx,Bartolo:2016ami,Sesana:2017vsj,Miller:2016krr,Nakamura:2016hna,Davoudiasl:2016mwf} including those for a GW background \cite{Cholis:2016xvo,Clesse:2016ajp,Wang:2016ana,Nakama:2016gzw,Crocker:2017agi,Kovetz:2017rvv,Garcia-Bellido:2017aan}. A bevy of other constraints (for a recent overview see \cite{Carr:2016drx}) on the existence and effects of PBH have been recently re-examined and new constraints proposed including those from dynamical effects in dwarf galaxies  \cite{Brandt:2016aco,Koushiappas:2017chw}, radio and X-ray sources \cite{Gaggero:2016dpq,Inoue:2017csr}, cosmic microwave background (CMB) measurements \cite{Chen:2016pud,Ali-Haimoud:2016mbv,Clark:2016nst,Georg:2017mqk,Poulin:2017bwe,Nakama:2017xvq}, ionization history \cite{Horowitz:2016lib}, quasar micro-lensing \cite{Mediavilla:2017bok}, neutron star capture \cite{Capela:2013yf}, lensing of radio bursts \cite{Munoz:2016tmg}, near infrared and cosmic infrared background \cite{Kashlinsky:2016sdv}, 21 cm measurements \cite{Gong:2017sie}, current and future pulsar timing arrays \cite{Inomata:2016rbd,Schutz:2016khr,Orlofsky:2016vbd},  lensing for intermediate mass PBH \cite{Axelrod:2016nkp}, future strong lensing tests \cite{Dai:2016igl}, orbital eccentricity determination \cite{Cholis:2016kqi}, spin distribution evaluation \cite{Chiba:2017rvs}, spatial clustering \cite{Garcia-Bellido:2017xvr} along with effects of astrophysical uncertainties on PBH constraints \cite{Green:2017qoa}.} A great deal of work has also been devoted to mechanisms for amplifying density perturbations toward the end of inflation to allow production of PBHs~\cite{Bartolo:2016ami,Kawasaki:2016ijp,Ezquiaga:2017fvi,Cheng:2016qzb,Kawasaki:2016pql,
Garcia-Bellido:2016dkw,Domcke:2017fix,Inomata:2016rbd,Orlofsky:2016vbd,Inomata:2017okj,Garcia-Bellido:2017mdw,Raidal}.

In this work, we study a scenario where PBHs are responsible for non-thermal production of DM, including the possibility of producing the entirety of the relic abundance. This scenario invokes a period of very early matter domination (VEMD) that leads to the formation of PBHs within an extended mass range. Evaporation of PBHs in the ensuing RD phase creates DM particles after thermal freeze-out or freeze-in but prior to BBN. Evading tight observational constraints for evaporation after BBN sets an upper bound $\simeq 2 \times 10^8$ g on the maximum mass $M_{{\rm{max}}}$ of PBHs thus formed. We show that the correct DM relic abundance can be obtained within the DM mass range $m_{{\rm{DM}}} = 100$ GeV-$10$ TeV, provided that the scalar power spectrum at small scales (relevant for PBH formation) is enhanced by a factor of ${\cal O}(10^5)$ relative to its value at scales probed by the cosmic microwave background (CMB) experiments. The upper limit on $M_{{\rm{max}}}$ implies that transition from VEMD to RD should occur when the Hubble expansion rate is $H_{{\rm{reh}}} \gtrsim {\cal O}(100)$ GeV. The observed DM abundance can be accommodated in cases when thermal freeze-out or freeze-in lead to underproduction of DM. The former case happens when DM annihilation rate is larger than the nominal value for the WIMP miracle, i.e., $\langle \sigma_{{\rm{ann}}} v \rangle_f > 3 \times 10^{-26}$ cm$^3$ s$^{-1}$. The latter case occurs when DM is extremely weakly coupled to the standard model (SM) particles resulting in a very small annihilation cross-section, possibly $\sigma_{{\rm{ann}}} \sim M^{-2}_{\rm{p}}$.          

We show that the required enhancement of the power spectrum is compatible with the Planck limits on the scalar spectral index and its running within 2$\sigma$ (for previous work on PBH in the context of a running spectral index, see \cite{Drees:2011hb,Drees:2011yz}). Such amplification is also attainable, for example, in models where the inflaton undergoes a brief period of ultra slow-roll motion toward the end of inflation.
%is also within the range that can be reached in the recently proposed models to amplify inflationary perturbations at small scales via ultra slow-roll motion.
We do not present an inflationary model that achieves this, however, as our goal in this work is to discuss the main ingredients for non-thermal DM production via evaporation of PBHs and identify the allowed parameter space. An explicit model that addresses inflation with the desired enhancement of the power spectrum as well as the origin of the VEMD era is the subject of a future work.       

The rest of this paper is organized as follows. In Section II, we briefly review PBH production in the early universe in both RD and (V)EMD phases (focusing on the latter). In Section III, we discuss non-thermal DM production via evaporation of PBHs. We present the main results in Secion IV. We close the paper by a brief dioscussion and conclusion in Section V. An Appendix includes some calculational details of our results presented in the main body of the paper.

\section{Formation of PBHs in very early matter domination}

PBHs are postulated to form from density fluctuations in the post-inflationary early universe. For the standard cosmology, the universe existed in a RD stage after the reheating process that followed inflation, and remained there until matter-radiation equality was reached in a post-BBN and pre-CMB era. However, as mentioned above, it is possible that there existed a period of (V)EMD that ended before the onset of BBN. 

It is well known that density fluctuations exhibit remarkably different growth behavior depending on the form of the dominant background energy component, be it radiation or matter. Here we briefly review the formation of PBHs in these two different background scenarios, and will then explore some of the consequences of PBH formation in the context of a VEMD scenario (for some of the works on EMD and PBH see, for example, \cite{Khlopov:1985jw,Polnarev:1986bi,Georg:2016yxa,Georg:2017mqk}).

In the case of a RD universe, a density fluctuation of ${\cal O}(1)$ would need to overcome the radiative pressure and thus would have a characteristic size on the order of the scale of the horizon. Assuming a Gaussian perturbation profile with root-mean-square amplitude $\delta(M)$, the fractional energy density of the universe that goes into PBHs with mass $M$ is given by~\cite{CKS}
\bea \label{betaR}
\beta(M) \approx K \delta^{2 \gamma}(M)\textrm{erfc}\left(\frac{\delta_c}{\sqrt{2}\delta(M)}\right),
\eea
where $\gamma \simeq 0.36$~\cite{NJ1,Musco1,Musco2,Musco3,KHA}, $K \simeq 3.3$~\cite{NJ2}, and $\delta_c \simeq 0.45$~\cite{Musco1,Musco2,Musco3} (see~\cite{HYK} for a smaller value of $\delta_c$). We note that after PBH formation $\beta$ increases $\propto a(t)$ during RD, where $a(t)$ is the scale factor.

However, the situation is altered if PBH formation takes place during a period of (V)EMD. In that case one arrives at \cite{Khlopov:1985jw,Polnarev:1986bi}
\bea \label{betatheory}
\beta(M) \approx 2\times 10^{-2}\delta^{13/2}(M) ,
\eea
where $\delta(M)$ denotes the amplitude of perturbations for a mode that eventually collapses to form a PBH with mass $M$ when it enters the horizon.
%, and $M$ is the mass contained within the horizon at that time. 
%PBHs with a mass $M$ are subsequently formed in this epoch once the mode whose wavelength encompasses a mass $M$ undergoes sufficient growth to %achieve a size of $\delta\sim{\cal O}(1)$~\cite{Khlopov:1985jw,Polnarev:1986bi} (provided that matter domination lasts long enough for this to happen). 
Such a perturbation enters the horizon when $H \simeq 4 \pi M^2_{{\rm{p}}}/M$ (where $M_{{\rm{p}}}$ is the Planck mass). 
%with an initial amplitude $\delta(M)$. 
The amplitude then grows according to $\delta \propto a \propto H^{-2/3}$ and black hole formation occurs at
\begin{equation} \label{formation}
H_{{\rm{form}}} \simeq {4 \pi M^2_{{\rm{p}}} \over M} \delta^{3/2}(M),
\end{equation}
when $\delta \sim {\cal O}(1)$. It is important to note that subhorizon fluctuations can form black holes due to the absence of pressure in this case. 
%In both Eqs.~(\ref{betaR},\ref{betatheory}), $\beta$ is the fractional energy density of PBHs at the time of formation. 
%In the RD phase $\beta$ increases in time $\propto a(t)$ (with $a$ being the scale factor), while it 
We also note that $\beta$ remains constant during the (V)EMD era.

The minimum mass of PBHs formed during the VEMD era, $M_{{\rm{min}}}$, depends on the details of the thermal history between the end of inflation (characterized by $H_{{\rm{inf}}}$) and the start of the VMED era (characterized by $H_0$), namely the window $H_0 \lesssim H \lesssim H_{{\rm{inf}}}$. An absolute lower bound on $M$ can be found by noticing that the minimal inflationary fluctuation wavelength is $\sim H^{-1}_{{\rm{inf}}}$, which implies that 
\be \label{Mmin}
M_{{\rm{min}}} \gtrsim {4 \pi M^2_{{\rm{p}}} \over H_{{\rm{inf}}}}. 
\ee
The maximum mass $M_{{\rm{max}}}$ corresponds to the mode whose amplitude reaches ${\cal O}(1)$ at the end of the VEMD epoch, which results in
\begin{equation} \label{reh}
M_{{\rm{max}}} \simeq {4 \pi M^2_{{\rm{p}}} \over H_{{\rm{reh}}}} \delta^{3/2}(M_{{\rm{max}}}) ,
\ee
where $H_{{\rm{reh}}}$ denotes the Hubble rate when the VEMD epoch ends and the universe enters the RD phase. In order to avoid very tight post-BBN constraints on evaporation of PBHs (for example, see~\cite{Carr:2009jm}), we require that all PBHs formed during the VEMD era evaporate before BBN. As we shall see, from Eq.~(\ref{lifetime}) in the following section, this results in an upper bound of $M_{{\rm{max}}} \lesssim 2 \times 10^8$ g. 

One comment is in order before moving to the next Section. Since transition from VEMD to RD is not instantaneous, one should not take the above expression for $M_{{\rm{max}}}$ as exact. The spectrum of PBHs formed during VEMD is not suddenly cut at $M_{{\rm{max}}}$. Instead, there is a quick drop in $\beta(M)$ around $M_{{\rm{max}}}$ signifying the transition from VEMD to RD. In fact, PBHs with a mass (much) larger than $M_{{\rm{max}}}$ may form in the following RD phase from the collapse of fluctuation modes that enter the hroizon then. However, the abundance of such PBHs is extremely suppressed due to its exponential dependence on $\delta(M)$ as seen in~(\ref{betaR}). Therefore, $M_{{\rm{max}}}$ provides a good approximation of the mass above which PBH formation during VEMD ceases to be important.

\section{Non-thermal DM from evaporation of PBHs}

Here we study a scenario in which the entire DM relic abundance is due to evaporation of PBHs formed during an epoch of VEMD. In passing, we note that PBHs formed in a RD phase can also produce DM particles via Hawking radiation. However, the exponential dependence of $\beta(M)$ on $\delta(M)$ in this case, see Eq.~(\ref{betaR}), implies that a parametrically larger $\delta(M)$ and a higher level of tuning are needed in this case in order to obtain the correct DM relic abundance. For this reason, we focus on DM production from PBHs formed in a VEMD phase.

PBHs with mass $M$ evaporate via Hawking radiation~\cite{Hawking} and have a lifetime
\begin{equation} \label{lifetime}
t_{{\rm{eva}}} = {80 M^3 \over \pi M^4_{{\rm{p}}}} ,
\end{equation}
giving rise to particles with a thermal spectrum at the Hawking temperature
\begin{equation} \label{Htemperature}
T_H = {M^2_{{\rm{p}}} \over M} .
\end{equation}

Evaporation of PBHs produces all particles that have a mass below their corresponding Hawking temperature. This implies that DM particles will also be produced as long as $m_{{\rm{DM}}} \ll T_H$~\cite{GreenDM}. For $M_{{\rm{max}}} \lesssim 2 \times 10^8$ g, this requires that $T_H \gtrsim 50$ TeV, and hence implying production of particles that are lighter than $\sim 50$ TeV\footnote{This inlcudes possible unwanted relics whose late decay may ruin the success of BBN, which leads to constrarints on $\beta(M)$~\cite{gravitino,moduli}.}. 
%Since all superparticles eventually (cascade) decay to the DM particle (i.e., the lightest superparticle), the branching fraction for producing DM from %evaporation of PBHs is ${\cal O}(1)$. 

PBHs with mass $M_{{\rm{max}}} \lesssim 2 \times 10^8$ g evaporate in the RD phase of the universe at a temperature $T_{{\rm{BBN}}} \approx$ 1 MeV. This late process can be responsible for the entire observed DM relic abundance in cases where thermal freeze-out or freeze-in lead to underproduction of DM. Thermal underproduction via freeze-out occurs for WIMPs with a large annihilaiton rate $\langle \sigma_{{\rm{ann}}} v \rangle_f > 3 \times 10^{-26}$ cm$^3$ s$^{-1}$.\footnote{DM particles produced from PBH evaporation will not undergo further annihilation if $\langle \sigma_{{\rm{ann}}} v \rangle < 3 \times 10^{-26}$ cm$^3$ s$^{-1}$ $(T_f/T_{{\rm{BBN}}})$, where $T_f \sim m_{{\rm{DM}}}/20$. This is the case for DM masses upto 10 TeV as indicated by the latest Fermi-LAT constraints~\cite{fermi2015,fermi2016}.} Thermal underproduction from freeze-in can happen if DM has extremely weak coupling to the SM particles, perhaps even gravitationally suppressed interactions resulting in a very small annihilation cross section $\sigma_{{\rm{ann}}} \sim M^{-2}_{p}$. 
%We therefore have a realization of the ``Branching" scenario of non-thermal DM~\cite{branching} where the DM relic density is entirely set by a late process %(PBH evaporation in this case).   

From the conservation of energy, and assuming that there is no other entropy generating process after the transition from VEMD to RD, evaporation of PBHs with mass $M$ results in a DM abundance
\be \label{DMdens1}
\left({n_{{\rm{DM}}} \over s}\right)_M \sim {Br}_{{\rm{DM}}} \left({n_{{\rm{PBH}}}(M) \over s}\right)  \left({M \over T_H}\right) = {Br}_{{\rm{DM}}}  \left({n_{{\rm{PBH}}}(M) \over s}\right) \left({M \over M_{{\rm{p}}}}\right)^2,
\ee
where $n_{{\rm{PBH}}}(M)$ and $s$ are the number density of PBHs and the entropy density in the RD phase, respectively, and ${Br}_{{\rm{DM}}}$ denotes the fraction of energy density in PBHs that goes into DM particles. For supersymmetric (SUSY) DM, we have ${Br}_{{\rm{DM}}} \sim 1$ in the case that all SUSY particles have a mass below $T_H$. The reason being that all SM particles and their SUSY partners are produced from PBH evaporation in this case with the latter eventually decaying to DM. However, ${Br}_{{\rm{DM}}} < 1$ if some of the SUSY particles have a mass above $T_H$. In the case that DM interacts extremely weakly with the SM paricles (and their SUSY partners), ${Br}_{{\rm{DM}}}$ can be as small as ${\cal O}(10^{-2})$ based on direct production of DM along with all SM degrees of freedom (and their SUSY partners) from PBH evaporation.       

The parameter $\beta(M)$ is related to the DM abundance through
\be \label{beta}
\beta(M) = \left({\rho_{{\rm{PBH}}}(M) \over \rho_{{\rm{tot}}}}\right)_{{\rm{reh}}} = {4 M \over 3 T_{{\rm{reh}}}} \left({n_{{\rm{PBH}}}(M) \over s}\right) ,
\ee
and hence
\be \label{DMdens2}
\left({n_{{\rm{DM}}} \over s}\right)_M \sim Br_{{\rm{DM}}} \beta(M) \left({3 T_{{\rm{reh}}} M \over 4 M^2_{{\rm{p}}}}\right). 
\ee
%
%Requiring that the contribution of PBHs with mass $M$ does not exceed 
The observed DM relic abundance is
\begin{equation} \label{DMdens3}
\left(\frac{n_{{\rm{DM}}}}{s}\right)_{{\rm{obs}}} \simeq 4 \times 10^{-12} \bigg(\frac{100\,GeV}{m_{{\rm{DM}}}}\bigg) .
\end{equation}
Requiring that the contribution from PBHs with mass $M$ does not exceed this value, and after using Eq.~(\ref{Htemperature},\ref{DMdens1},\ref{beta},\ref{DMdens2})), we arrive at the following relation
%~\cite{Carr:2009jm}
%
\be  \label{CKSYbetaprimeconstraint}
\beta(M) \simeq 2.3 \times 10^{-26} Br^{-1}_{{\rm{DM}}} \bigg(\frac{g_{*,{\rm{reh}}}}{106.75}\bigg)^{1/4} \bigg(\frac{10^{11} ~ {\rm g}}{M}\bigg) \bigg(\frac{100 ~ {\rm GeV}}{m_{{\rm{DM}}}}\bigg) \bigg(\frac{M_{{\rm{p}}}}{H_{{\rm{reh}}}}\bigg)^{1/2},
\ee
where $g_{*,{\rm{reh}}}$ denotes the number of relativistic degrees of freedom at $T_{{\rm{reh}}}$. We note that the similar expression in~\cite{Carr:2009jm}, which holds in the case of PBH formation in the RD phase, includes formation temperature $T_i$ instead of the reheat temperature $T_{{\rm{reh}}}$.  

In reality, all of the PBHs formed within an extended mass range during the VEMD epoch contribute to the DM relic density. However, see Eq.~(\ref{CKSYbetaprimeconstraint}), the constraint on $\beta(M)$ becomes weaker for lighter black holes. To be precise, one has to integrate over the whole relevant mass range relevant to find the total contribution to the DM abundance. As shown in the Appendix, that integral is typically dominated by the heaviest PBHs in the mass range. We thus have 
%Hence, requiring that the entire DM relic abundance is due to evaporation of PBHs results in
%
\be \label{betaobservation}
\beta(M_{{\rm{max}}}) \simeq 10^{-23} Br^{-1}_{{\rm{DM}}}\bigg(\frac{g_{*,{\rm{reh}}}}{106.75}\bigg)^{1/4} \left({2 \times 10^8 ~ {\rm g} \over M_{{\rm{max}}}}\right) \bigg(\frac{100 ~ {\rm GeV}}{m_{{\rm{DM}}}}\bigg) \bigg(\frac{M_{{\rm{p}}}}{H_{{\rm{reh}}}}\bigg)^{1/2},
\ee
where $M_{{\rm{max}}}$ is normalized to its largest value for evaporatation before the onset of BBN. By using Eq.~(\ref{reh}), we can cast this expression in terms of $M_{{\rm{max}}}$ and $\delta(M_{{\rm{max}}})$    
\be \label{betaobservation2}
\beta(M_{{\rm{max}}}) \simeq 9 \times 10^{-24} Br^{-1}_{{\rm{DM}}} \bigg(\frac{g_{*,{\rm{reh}}}}{106.75}\bigg)^{1/4} \left({2 \times 10^8 ~ {\rm g} \over M_{{\rm{max}}}}\right)^{1/2} \bigg(\frac{100 ~ {\rm GeV}}{m_{{\rm{DM}}}}\bigg) \left(\delta(M_{{\rm{max}}})\right)^{-3/4}.
\ee

\section{Results}

Equating the theoretical prediction for $\beta(M_{{\rm{max}}})$ in Eq.~(\ref{betatheory}) with that satisfying the observational constraint in Eq.~(\ref{betaobservation2}) singles out the value of $\delta(M_{{\rm{max}}})$ that is required to obtain the correct DM relic abundance for a given value of $m_{{\rm{DM}}}$
\be \label{deltaDM}
\left(\delta(M_{{\rm{max}}})\right)^{29/4} \simeq 5 \times 10^{-22} Br^{-1}_{{\rm{DM}}} \bigg(\frac{g_{*,{\rm{reh}}}}{106.75}\bigg)^{1/4} \left({2 \times 10^8 ~ {\rm g} \over M_{{\rm{max}}}}\right)^{1/2} \bigg(\frac{100 ~ {\rm GeV}}{m_{{\rm{DM}}}}\bigg).
\ee
We can trade out $\delta$ for the amplitude of the scalar power spectrum $A_s \equiv 25 \delta^2/4$~\cite{Silkrunning}. Planck has measured a value $A_s = 2.196 \times 10^{-9}$ at the pivot scale of $k_* = 0.05$ Mpc$^{-1}$~\cite{planck}. We translate the value of $\delta(M_{{\rm{max}}})$ in Eq.~(\ref{deltaDM}) to the enhancement factor in $A_s$, denoted by $f$, from $k_*$ to that relevant for formaing PBHs with mass $M_{{\rm{max}}}$. 

Fig.~1 depicts the $\beta(M_{{\rm{max}}})$ curve from theory and bands representing the observational constraint as a function of $f$ for $M_{{\rm{max}}} = 2 \times 10^8$ g. In the left panel, $Br_{{\rm{DM}}} = 1$ and the band corresponds to $m_{{\rm{DM}}} =$100 GeV-10 TeV mass range. In the right panel, $m_{{\rm{DM}}} = 1$ TeV and the band corresponds to $Br_{{\rm{DM}}} = 10^{-2}-1$ range. The intersection region lies between $f \approx 6 \times 10^4-2 \times 10^5$ and $f \approx (1-4) \times 10^5$ in the left and right panels respectively.
%One sees a sweet spot $6\times10^4 < f < 2\times10^5$ where the curve and the band intersect. 

After using Eqs.~(\ref{betatheory},\ref{reh},\ref{betaobservation}), we find the following expression for $H_{{\rm{reh}}}$ 
%at which the epoch of VEMD ends
%
\be \label{end} 
H_{{\rm{reh}}} \approx (200 ~ {\rm GeV}) \times Br^{-6/29}_{{\rm{DM}}} \bigg(\frac{g_{*,{\rm{reh}}}}{106.75}\bigg)^{3/58} \bigg(\frac{100~ {\rm GeV}}{m_{{\rm{DM}}}}\bigg)^{6/29} \bigg(\frac{2\times10^{8}g}{M_{{\rm{max}}}}\bigg)^{32/29}.
\ee
This results in $H_{{\rm{reh}}} \approx (200 - 500)$ GeV within the DM mass range $m_{{\rm{DM}}} = $100 GeV-10 TeV, for $M_{{\rm{max}}} = 2 \times 10^8$ g, with a very mild dependence on $Br_{{\rm{DM}}}$. This corresponds to a very high reheat temperature $T_{{\rm{reh}}} \sim 10^{10}$ GeV, assuming that the universe instantly thermalizes, which is why we dubbed the era as ``VEMD". {\bf Considering that fluctuations grow as $\delta \propto a \propto H^{-2/3}$ in this epoch, formation of PBHs with mass $M_{\rm max} = 2 \times 10^8$ g from perturbations whose initial amplitude is enhanced according to Fig.~1 requires that VEMD starts no later than $H_0 \approx 10^6$ GeV. This sets an absolute lower bound $H_{\rm inf} \gtrsim 10^6$ GeV, which essentially excludes models of low scale inflation.}  

We see from Eq.~(\ref{deltaDM}) that a larger value of $\delta(M_{{\rm{max}}})$, and hence $f$, is needed when $M_{{\rm{max}}} < 2 \times 10^8$ g. Also, Eq.~(\ref{end}) implies a larger $H_{{\rm{reh}}}$ in this case. Therefore the scenario will be least constrained for $M_{{\rm{max}}} \simeq 2 \times 10^8$ g.

\begin{figure}
    \centering
    \includegraphics[width=1\linewidth]{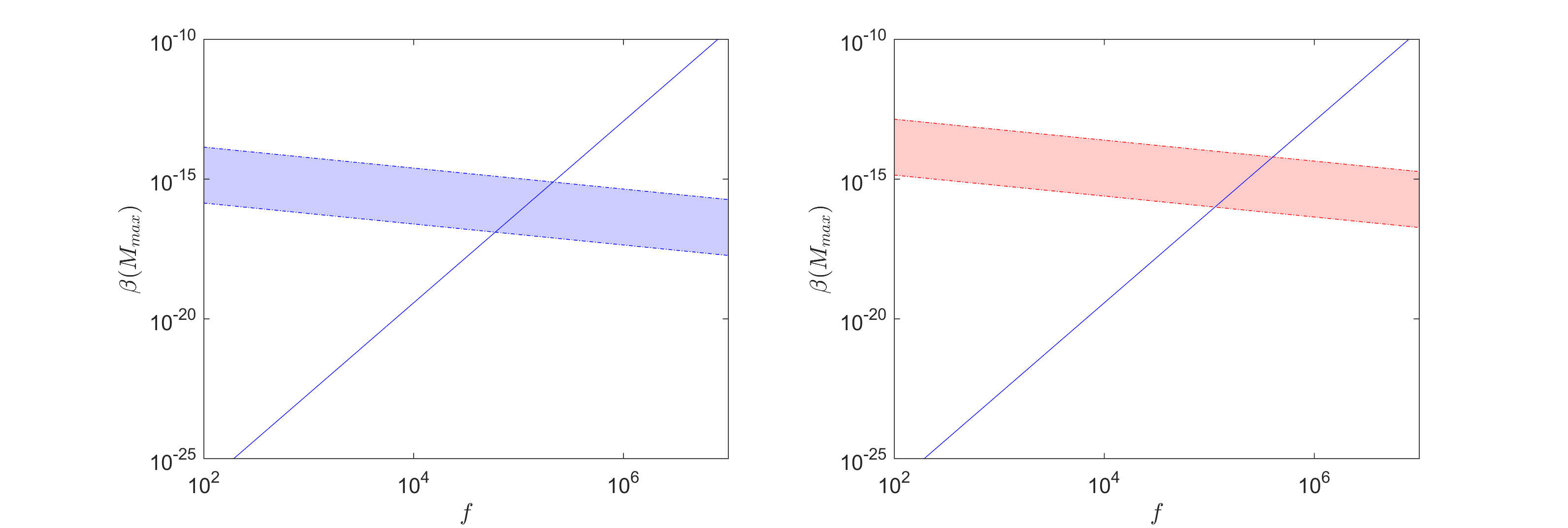}
    \caption{Left: the solid line corresponds to the theoretical expression for $\beta(M_{{\rm{max}}})$~(\ref{betatheory}) as a function of the enhancement factor $f$ in the scalar power spectrum at scales relevant for PBH formation. The shaded band shows the observational constraint on $\beta(M_{{\rm{max}}})$ in Eq.~(\ref{betaobservation2}) for $Br_{{\rm{DM}}} = 1$ and $m_{{\rm{DM}}} =$ 100 GeV - 10 TeV.
% A larger DM mass extends this band toward smaller values of $\beta$. 
%The solid line corresponds to the theoretical expression for $\beta(M_{{\rm{max}}})$ expressed in terms of the enhancement factor $f$. 
%The intersection region lies between $f \approx 6\times10^4 - 2\times10^5$. 
Right: same as the left panel, but the shaded band shows the $\beta(M_{{\rm{max}}})$ constraint for $m_{{\rm{DM}}} =$ 1 TeV and $Br_{{\rm{DM}}} =10^{-2} - 1$ range. We have taken $M_{{\rm{max}}} = 2 \times 10^8$ g for both panels.
%, with larger branching extending lower. Here, the intersection lies between $f \approx 1\times10^5 - 4\times10^5$
}
    \label{fig:beta}
\end{figure}

One question that arises is whether the large enhancement of the power spectrum that is needed at small scales $f \sim {\cal O}(10^5)$ 
%to form a sufficient number of PBHs 
is compatible with Planck limits on $n_s$ and its running at the pivot scale $k_*$. Following~\cite{Munoz:2016tmg}, we can write
\be \label{running}
{\rm ln} A_s(k) = {\rm ln} A_s(k_*) + (n_s -1) {\rm ln} \left({k \over k_*}\right) + {1 \over 2} \alpha_s {\rm ln}^2 \left({k \over k_*}\right) + {1 \over 6} \beta_s {\rm ln}^3 \left({k \over k_*}\right) ,
\ee
where $n_s = 0.9655$ at $k_*$. Choosing $k = k_{{\rm{max}}}$, where $k_{{\rm{max}}}$ denotes the mode that eventually collapses to a PBHs with mass $M_{{\rm{max}}}$, we have
\be \label{running2}
{\rm ln} f =  (n_s -1) {\rm ln} \left({k_{{\rm{max}}} \over k_*}\right) + {1 \over 2} \alpha_s {\rm ln}^2 \left({k_{{\rm{max}}} \over k_*}\right) + {1 \over 6} \beta_s {\rm ln}^3 \left({k_{{\rm{max}}} \over k_*}\right) .
\ee
The question is now whether values of $f$ inferred from Fig.~\ref{fig:beta} are compatible with constraints from Planck data on the running parameters $\alpha_s$ and $\beta_s$ at the pivot scale. As shown in the Appendix
\be \label{running3}
{\rm ln} \left({k_{{\rm{max}}} \over k_*}\right) \approx 47.7 + {1 \over 4} {\rm ln} \delta(M_{{\rm{max}}}) .
\ee

Using this expression, and the relation $f = (\delta(M_{{\rm{max}}})/\delta_*)^2$, we can now check the consistency of Eq.~(\ref{running2}) with Planck data. In Fig.~\ref{fig:planck}, we show a band in the $\alpha_s-\beta_s$ plane that corresponds to the intersection region in the left panel of Fig.~1 through Eq.~(\ref{running3}). This band is in agreement with Planck constraints on $\alpha_s$ and $\beta_s$ at the 2$\sigma$ level.

\begin{figure}
    \centering
    \includegraphics[width=.7\linewidth]{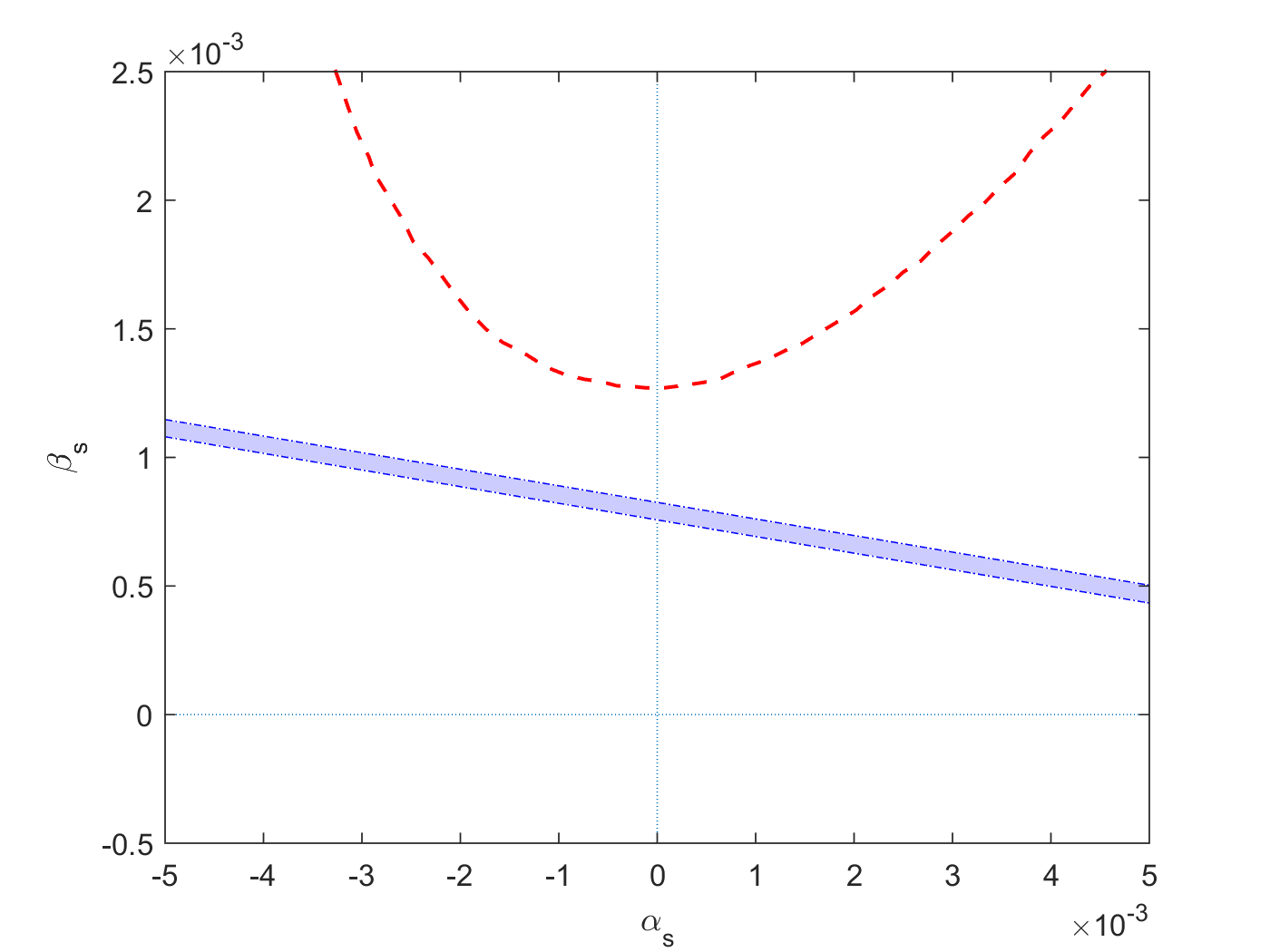}
    \caption{The shaded band in the $\alpha_s-\beta_s$ plane corresponding to the intersection region in the left panel of Fig.~\ref{fig:beta}. 
%for $m_{{\rm{DM}}} =$ 100 GeV - 10 TeV, shown in the $\alpha_s-\beta_s$ plane. 
The 68$\%$ confidence ellipse from Planck 2015~\cite{Silkrunning} on the running of the scalar spectral index is outlined by the red dashed line. }
    \label{fig:planck}
\end{figure}

Another question that naturally arises concerns the mechanism behind a large enhancement of the power spectrum for modes around $k_{{\rm{max}}}$. It is known in models of hybrid inflation that large density perturbations, which could lead to formation of PBHs, can be obtained toward the end of inflation~\cite{Juanhybrid}. This is also possible to achieve via multiple phases of inflation in single field models~\cite{Raidal}, or by a brief period of ultra slow-roll motion toward the end of inflation~\cite{Garcia-Bellido:2017mdw,JuanHiggs,GP,HM,Gong,BT}. It has been shown~\cite{HM} that an amplification of the power spectrum up to a factor of $10^7$ in 10 e-folds (or more) can be obtained from ultra slow-roll inflation near an inflection point. This fits well with the requirement in our scenario, namely an enhancement factor $f \sim{\cal O}(10^5)$ within a few e-folds including the mode $k_{{\rm{max}}}$. While an explicit model to achieve this is beyond the scope of this paper, it is assuring that the desirable enhancement in the power spectrum is both compatible with the Planck limits and achievable in models of single field inflation.

\section{Discussion and Conclusion}

We now turn to a discussion of the possible origin and consequences of a VEMD phase, formation of PBHs in the subsequent RD phase, and possible issues related to very light PBHs. Any implemenation of our scenario within a specific model must be aware of these issues along with possible ways of addressing them.
\vskip 1.5mm
\noindent
{\bf Origin and consequences of VEMD-}
% We have not gone into details of VEMD in our model-independent analysis. 
An era of VEMD can arise from oscillations of a very heavy modulus field that dominates the energy density of the universe soon after the end of inflation. It may also start right at the end of inflation when the universe is dominated by inflaton oscillations that eventually decay slowly via perturbative channels (for example, when the inflaton has gravitationally suppressed couplings to the visible sector fields). It may also be possible that initial stages of inflaton decay occurs via non-perturbative effects~\cite{STB,KLS1,KLS2} (for reviews, see~\cite{Review1,Review2}), and the zero-mode quanta of the inflaton (or other scalar fields produced during this process) come to dominate the universe at some point. 

One might worry about dangerous consequences of high reheat temperatures $T_{{\rm{reh}}} \sim 10^{10}$ GeV at the end of the VEMD epoch (see Eq.~(\ref{end}) and the disucssion below it). Notably, a concern arises regarding thermal overproduction of gravitinos that decay after BBN~\cite{EKN,KL,BBB}. This can be avoided if the gravitino mass is $m_{3/2} \gtrsim 50$ TeV so that gravitinos decay before the onset of BBN. Thermal gravitino production will be totally irrelevant if $m_{3/2} \gg 10^{10}$ GeV, which can happen in some string theory cnstructions (for example, see~\cite{LVS}). Gravitino production can also be suppressed if the universe has the same equation of state as radiation but thermalziation is delayed and full thermal equilibrium is not established at the end of the VEMD \cite{AM}. 
\vskip 1.5mm
\noindent
{\bf PBH formation in the RD phase-} The relation for $H_{{\rm{reh}}}$ in Eq.~(\ref{end}) ensures that PBHs whose mass ia larfger than $M_{{\rm{max}}}$ are practically not produced in the VEMD epoch. However, enhancement of the power spectrum for modes around $k_{{\rm{max}}}$ may result in the formation of heavier PBHs in the ensuing RD phase. This can happen from the collapse of those modes that enter the horizon when $H \lesssim H_{{\rm{reh}}}$ (recall that in a RD universe pressure dominates over gravity for subhorizon modes thus preventing their collapse).  We do not expect formation of PBHs with exceedingly large masses in the RD phase as they would correspond to modes with $k \ll k_{{\rm{max}}}$, for which $\delta(M) \ll \delta(M_{{\rm{max}}})$ due to the rapid fall off in the power spectrum far away from $k_{{\rm{max}}}$. In fact, because of the exponential dependence of $\beta(M)$ on $\delta(M)$ in the RD phase, even moderate suppression of the power spectrum at $k < k_{{\rm{max}}}$ can yield a substantial decrease in $\beta(M)$ in accordance with the most stringent observational limits. The abundance of PBHs that may form during RD follows from Eq.~(\ref{betaR}) by setting $\delta = \delta(M_{{\rm{max}}})$. For values of $\delta(M_{{\rm{max}}})$ corresponding to te intersection regions in Fig.~1, we have checked that $\beta(M) << 10^{-30}$ for $M \gtrsim 2 \times 10^8$ g. This easily satisfies even the tightest observational constraints on the abundance of PBHs over the entire mass range that evaporate after BBN~\cite{Carr:2009jm}.
\vskip 1.5mm
\noindent
{\bf Effects of light PBHs-} Very light PBHs with mass $M \ll M_{{\rm{max}}}$ may form in the VEMD phase. As mentioned above, and shown in the Appendix, the contribution of such black holes to the DM relic abundance is typically negligible. However, they can act as a site for bubble nucleation in a first order phase transition and seed vacuum decay. This effect can be relevant for the electroweak vacuum that becomes unstable for a certain range of the top quark mass~\cite{Burda1}. It has been shown that PBHs that have a mass in the $(10^5-10^9) M_{{\rm{p}}}$ range can seed decay of the Higgs vacuum as the decay rate dominates over the Hawking evaporation~\cite{Burda2,Espinosa:2017sgp}. 

%This effect can be avoided if formation of PBHs with a mass $M \lesssim 10^9 M_{{\rm{p}}}$ during the VEMD epoch is prevented. 
The simplest possibility to avoid this effect is to have an inflationary scale that corresponds to $H_{{\rm{inf}}} \lesssim 10^{-9} M_{{\rm{p}}}$. Since inflation generates density perturbations with physical wavenumbers $k < H_{{\rm{inf}}}$, the condition $H_{{\rm{inf}}} \lesssim 10^{-9} M_{{\rm{p}}}$ ensures the absence of fluctuation modes that could collapse to form dangerous PBHs with mass $M \lesssim 10^9 M_{{\rm{p}}}$. Another possibility is to
%relax the restriction on $H_{{\rm{inf}}}$ but 
have a situation where $\delta(M) \ll \delta(M_{{\rm{max}}})$ for $M \lesssim 10^9 M_{{\rm{p}}}$ as a result of the rapid fall off in the power spectrum far from the mode $k_{{\rm{max}}}$. Formation of dangerously light PBHs can then be prevented if the corresponding fluctuations do not grow to become ${\cal O}(1)$ by the time when $H \simeq H_{{\rm{reh}}}$, without any restrictions on $H_{{\rm{inf}}}$.         
\vskip 1.5mm
\noindent

In conclusion, we have demonstrated that evaporation of PBHs can produce the entire DM relic abundance within the DM mass range of 100 GeV-10 TeV in cases with thermal underproduction. The necessary ingredient for this non-thermal scenario to be viable is to produce a sufficient abundance of PBHs whose  mass is below $\simeq 2 \times 10^8$ g, so that they evaporate before BBN. We have found that an epoch of VEMD can accommodate this if the scalar power spectrum at small scales is enhanced by a factor of ${\cal O}(10^5)$ relative to its value for the CMB modes. An explicit model that leads to inflation with the desired enhancement of the power spectrum as well as an epoch of VEMD will be the subject of a future investigation. 

\section{Acknowledgements}

The work of R.A. and J.O. is supported in part by NSF Grant No. PHY-1720174. J.B.D would like to acknowledge support from the Mitchell Institute for Fundamental Physics and Astronomy at Texas A\&M University. The authors acknowledge discussions with N. Afshordi and M. M. Sheikh-Jabbari. The authors also thank Andrew Powell, who worked on an earlier stage of this project.

\section{Appendix}
\subsection{Number of relevant e-foldings}
%We want to calculate the number of e-folds between the exit (during inflation) of the pivot scale and the mode associated with the PBH that evaporates at %BBN. 

For the standard thermal history, the number of e-foldings of inflation between the time when the pivot scale $k_* = 0.05$ Mpc$^{-1}$ left the horizon and the end of inflation is given by~\cite{LL,RG2015}
\begin{equation} \label{Nk*1}
N_{k_*} \approx 63.5 + {1 \over 4} {\rm ln} {3 H^2_{{\rm{inf}}} \over (8 \pi M_{{\rm{p}}})^2} + {1 \over 6} {H_{{\rm{R}}} \over H_{{\rm{inf}}}}. 
\end{equation}
Here $H_{{\rm{inf}}}$ and $H_{{\rm{R}}}$ denote the Hubble rate at the end of inflation and when reheating after inflation completes respectively, assuming that the universe has the same equation of state as a MD phase for $H_{{\rm{R}}} < H < H_{{\rm{inf}}}$. In the presence of an epoch of VEMD for $H_{{\rm{reh}}} < H < H_0$, this relation is modified as follows
\be \label{Nk*2}
N_{k_*} \approx 63.5 + {1 \over 4} {\rm ln} {3 H^2_{{\rm{inf}}} \over (8 \pi M_{{\rm{p}}})^2} + {1 \over 6} {H_{{\rm{R}}} \over H_{{\rm{inf}}}} + {1 \over 6} {H_{{\rm{reh}}} \over H_0}. 
\ee

The number of e-foldings relevant for the mode $k_0$ that enters the horizon at $H = H_0$ is given by 
\begin{equation} \label{Nk0}
N_0 = {1 \over 3} {\rm ln} {H_{{\rm{inf}}} \over H_{{\rm{R}}}} + \frac{1}{2} \ln \frac{H_{R}}{H_0}.
\end{equation}
The first and second terms on the right-hand side of the equation take evolution in the MD phase between $H_{{\rm{inf}}}$ and $H_{{\rm{R}}}$ and the RD phase between $H_{{\rm{R}}}$ and $H_0$, respectively, into account. The number of e-foldings relevant for the mode $k_{{\rm{max}}}$ that eventually collapses to form PBHs with mass $M_{{\rm{max}}}$ follows from
\be \label{Nkmax}    
N_{k_{{\rm{max}}}} - N_0 = {1 \over 3} {\rm ln} {H_0 \over H_{{\rm{max}}}} ,
\ee
where $H_{{\rm{max}}} = 4 \pi M^2_{{\rm{p}}}/M_{{\rm{max}}}$ is the Hubble rate when the mode $k_{{\rm{max}}}$ enters the horizon.  

After using Eqs.~(\ref{Nk*2},\ref{Nk0},\ref{Nkmax}), and with the help of~(\ref{reh}), we find
\begin{equation} 
N_{k_*} - N_{k_{{\rm{max}}}} \approx 47.7 + \frac{1}{4}\ln\delta(M_{{\rm{max}}}).
\end{equation}
It is interesting to note that $H_{{\rm{inf}}}$ does not appear in this expression. However, it implicitly enters as we must have $H_0 \leq H_{{\rm{R}}} \leq H_{{\rm{inf}}}$. 
For the shaded band shown Fig.~2, we get $N_{k_*} - N_{M_{{\rm{max}}}} \approx 46.4-46.5$. 
%for \(\delta \approx 0.0046\), and \(N_{k_*} - N_{{\rm{PBH}}} \approx 46.5\) for \(\delta \approx 0.0087\).

\subsection{Integrating over PBHs in an extended mass range} \label{appendix_integral}

Here we derive the total contribution from evaporation of PBHs within a mass range $M_{{\rm{min}}} \leq M \leq M_{{\rm{max}}}$ to the DM relic abundance. For simplicity, we assume that $Br_{{\rm{DM}}} = 1$ in this derivation, but including $Br_{{\rm{DM}}}$ in the claculation is straightforward. In general, the number density of regions with mass $M$ in the early universe can be written as 
\be \label{NM}
n(M) = {\rho_{{\rm{tot}}} \over M},
\ee
where $\rho_{{\rm{tot}}}$ is the total energy density of the universe. Then the number density of PBHs whose mass is between $M$ and $M + dM$ follows from
\be \label{dNPBH1} 
dn_{{\rm{PBH}}}(M) = \beta(M) dn(M).
\ee
Since $\rho_{{\rm{tot}}}$ does not depend on $M$, we have $dn(M) M = - n(M) dM$. This results in
\be \label{dNPBH2}
dn_{{\rm{PBH}}}(M) =  {\beta(M) n(M) \over M} dM = \rho_{{\rm{tot}}} {\beta(M) \over M^2} dM,
\ee
and hence
\be \label{rhoPBH}
{\rho_{{\rm{PBH}}} \over \rho_{{\rm{tot}}}} = {\int_{M_{{\rm{min}}}}^{M_{{\rm{max}}}}{M dn_{{\rm{PBH}}}(M)} \over \rho_{{\rm{tot}}}} = \int_{M_{{\rm{min}}}}^{M_{{\rm{max}}}}{{\beta{M} \over M} dM}.
\ee
Also, after using the relation $\rho_{{\rm{tot}}}/s = 3T_{{\rm{reh}}}/4$ at the end of the VEMD epoch, we find:
\be \label{dNPBH3}
{dn_{{\rm{PBH}}}(M) \over s} =  {3 \over 4} T_{{\rm{reh}}} {\beta(M) \over M^2} dM .
\ee
% 
%where we have used $\rho_{{\rm{tot}}}/s = 3T_{{\rm{reh}}}/4$ at the end of the VEMD epoch. 
The corresponding contribution to the DM relic abundance, see Eq.~(\ref{DMdens1}), is given by
\be \label{dNDM1}
{dn_{{\rm{DM}}} \over s} \sim {dn_{{\rm{PBH}}}(M) \over s} {M^2 \over M^2_{{\rm{p}}}}.
\ee
After integrating over the entire mass range, we derive the analogue of Eq.~(\ref{DMdens2})
\be \label{NDM}
{n_{{\rm{DM}}} \over s} \sim \int_{M_{{\rm{min}}}}^{M_{{\rm{max}}}}{{3 T_{{\rm{reh}}} \over 4 M^2_{{\rm{p}}}} \beta(M) dM},
\ee
which generalizes Eq.(\ref{CKSYbetaprimeconstraint}) to: 
\begin{equation} \label{CKSYbetaprimeconstraint2}
\int_{M_{{\rm{min}}}}^{M_{{\rm{max}}}} \beta(M) \bigg(\frac{H_{{\rm{reh}}}}{M_{{\rm{p}}}}\bigg)^{1/2} \bigg(\frac{106.75}{g_{*,{\rm{reh}}}}\bigg)^{1/4} \frac{dM}{2 \times 10^8 ~ {\rm g}} \simeq 10^{-23} \bigg(\frac{100\,{\rm{GeV}}}{m_{{\rm{DM}}}}\bigg),
\end{equation}
where $\beta(M)$ is the theoretical prediction given in Eq.~(\ref{betatheory}). 

If $\beta(M)$ varies slowly in the mass range $M_{{\rm{min}}} \leq M \leq M_{{\rm{max}}}$, the above integral is $\propto M_{{\rm{max}}}$ and Eq.~(\ref{CKSYbetaprimeconstraint2}) reproduces the bound in (\ref{betaobservation}). The same conclusion holds as long as the minimum of $\beta(M)$ does not happen around $M_{{\rm{max}}}$. We expect that this be the case if the range of modes over which the power spectrum is enhancedinlcudes $k_{{\rm{max}}}$.   
%is not a decreasing  integral is dominated by the contribution he same conclusion holds as long as $\beta(M)$ is not a decreasing function of $M$ within the %given mass range.       

\bibliography{Final}

\end{document}